# Thinking Hard About Physics Before Calculating: An Example from Pendulum Physics

Zhiwei Chong 崇志伟

Malvern College at Qingdao, Shandong, China

This paper aims to show how to guide students with a familiar example to extract as much physics as possible before jumping into mathematical calculation. The period for a physical pendulum made up of a uniform rod is changed by attaching a piece of putty on it. The period for the combined system depends on the location of the putty. Simple reasoning without calculation shows that there are two locations for the putty that do not change the period of the physical pendulum: the axis and the center of percussion. Moreover, without calculation, we reason that there is at least one minimal period when the putty lies somewhere between these two locations. The underlying physics is that the period depends on the interplay between two factors: the additional torque due to putty and the associated moment of inertia, both of which depend on the location of the putty. They conflict with each other; a smaller (larger) moment of inertia is accompanied by a smaller (larger) torque. The minimal value of the period and the corresponding location for the putty can only be obtained by calculation.

What motivates me to write this article is the book *Fly by Night Physics* by Anthony Zee who also emphasizes thinking by quoting John Wheeler in the preface to his book [1]: Never never calculate unless you already know the answer! Another input is the qualitative analysis of simple harmonic motion, where the sinusoidal shape of the velocity versus time curve is not obtained by calculation but based purely on physical reasoning instead [2].

## Preliminary Considerations

The journey starts from the following two basic facts. For small amplitude oscillations, the period for a simple pendulum with length $\ell$ and the period for a physical pendulum made up of a uniform rod with length $L$ are

$$T_s = 2\pi\sqrt{\frac{\ell}{g}}, \quad T_p = 2\pi\sqrt{\frac{2L}{3g}}, \tag{1}$$

where the subscripts $s$ and $p$ stand for simple and physical, respectively. Now we put a simple pendulum and a physical pendulum side by side as illustrated in Fig. 1(a). If the length of the simple pendulum $\ell$ is $2L/3$, then the two pendulums oscillate with the same period. In particular, if they are released from rest at the same small angle, then the rod of the physical pendulum and the string of the simple pendulum keep parallel to each other as they oscillate. When $\ell \neq 2L/3$, say, $\ell < 2L/3$, then $T_s < T_p$, and the simple pendulum oscillates more briskly than the physical one. On the other hand, if $\ell > 2L/3$, then the simple pendulum oscillates more drowsily. These elementary facts help to understand the following experiment.

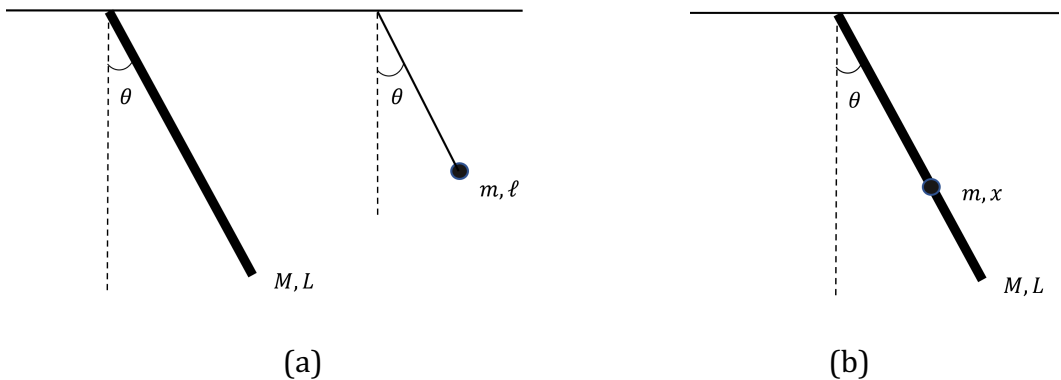

(a)          (b)

Fig. 1. (a) A physical pendulum made up of a unform rod with length $L$ is put side by side with a simple pendulum with string length $\ell = 2L/3$. (b) A piece of putty with mass $m$ is attached on the rod at distance $x$ from the axis.

Instead of putting a simple pendulum side by side with a physical pendulum, now we attach a piece of putty with mass $m$ on the rod of the physical pendulum at distance $x$ from the axis, as illustrated in Fig. 1 (b). Three questions were raised to my students:

1. Will the period change if the putty is attached at $x = 0$ or $x_p = 2L/3$?
2. How will the period change if the putty is attached at distance $0 < x < x_p$?
3. How will the period change if the putty is attached at distance $x > x_p$?

With the experience in the first experiment, students figured out rather easily that the period of the combined system remains the same as $T_p$ in the first question [3]. The period is less than $T_p$ in the second question, and greater in the third.

## Further Considerations

The following question becomes a bit more mathematical than physical: What's the shape of the graph for period $T$ against distance $x$? The answers to the three questions in the previous section provide key elements for the graph: (i) $T = T_p$ for $x = 0, x_p$, as indicated by points A and B in Fig. 2, (ii) $T > T_p$ for $x > x_p$, as indicated by point C, and (iii) $T < T_p$ for $x < x_p$, that is, the curve between A and B is below the line AB. What's troubling is the shape of the curve connecting A and B. In particular, how many extreme points between A and B? Without calculation, we resort to Occam's razor for the moment [4]. It is reasonable to consider the simplest possible case: the curve is convex up with only one minimum indicated by D in Fig. 2.

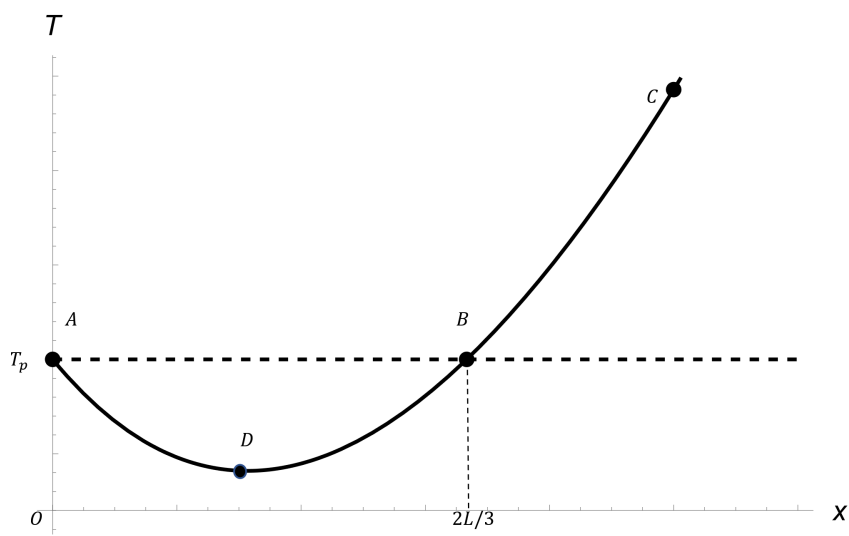

Fig. 2 Rough shape of the curve for period versus distance.

Before we start to determine mathematically the location that gives the minimal period, we want to understand its existence by physical considerations. The period is determined by two factors: one is the torque, and the other is the moment of inertia. The larger the torque and the smaller the moment of inertia are, the shorter the period is.

However, they cannot be achieved simultaneously in our problem. On one hand, the closer the putty is to the axis, the smaller the moment of inertia, which helps to result in a short period. However, the torque is smaller too, which does not help to shorten the period. When the putty is located further away from the axis, on the other hand, the discussion can be carried out in a similar way.

The above discussion can be further understood in the following way. The torque produced by putty is $mgx \sin \theta$, which is proportional to $x$. Its associated moment of inertia is $mx^2$, which is proportional to $x^2$. For a small value of $x$, the torque effect dominates the inertia effect. In other words, as $x$ starts increasing from 0, the period becomes smaller. However, this trend won't continue forever since the inertia effect dominates the torque effect as $x$ becomes large; the period stops decreasing and starts increasing at a certain value of $x$. As a result, there must exist a minimal value of period when the location of the putty nicely balances these two factors. An alternative explanation from the perspective of the change in the center of percussion due to the putty is provided in [5].

Furthermore, as $x$ becomes even larger, the inertia effect keeps dominating the torque effect, which actually makes the minimal period be the unique minimum. Up to this point we are quite confident that we have a rather good understanding of this problem without mathematical calculations. We are now at exactly the point "you already know the answer" by John Wheeler. It is time to calculate; the location of the putty for the minimum period can only be determined by calculation in the following section.

## Calculations

With the experience in solving the physical pendulum, my students were able to solve the problem by themselves. Denote the mass of the rod as $M$, then the torque with respect to the axis is

$$\tau = \frac{1}{2} MgL \sin \theta + mgx \sin \theta, \qquad (2)$$

where the first term is the torque generated by the rod's gravity and the second by putty. On the other hand, the total moment of inertia with respect to the axis is

$$I = \frac{1}{3}ML^2 + mx^2. \tag{3}$$

Making the small angle approximation $\sin\theta \approx \theta$ and applying Newton's second law for rotation give

$$\left(\frac{1}{3}ML^2 + mx^2\right)\ddot\theta + \left(\frac{1}{2}ML + mx\right)g\theta = 0. \tag{4}$$

The period is thereby

$$T = 2\pi\sqrt{\frac{\frac{1}{3}ML^2 + mx^2}{\left(\frac{1}{2}ML + mx\right)g}} = 2\pi\sqrt{\frac{2}{3g}\frac{ML^2 + 3mx^2}{ML + 2mx}}. \tag{5}$$

It is convenient to introduce two dimensionless variables, following Ref. [6],

$$r = \frac{M}{m}, \quad \tilde{x} = \frac{x}{L}. \tag{6}$$

Then the period can be rewritten as

$$T = 2\pi\sqrt{\frac{2L}{3g}f(\tilde{x})} = T_p\sqrt{f(\tilde{x})}, \quad f(\tilde{x}) = \frac{r + 3\tilde{x}^2}{r + 2\tilde{x}}. \tag{7}$$

The focus of our study is the function $f(\tilde{x})$, which is plotted in Fig. 3 with $r = 5$.

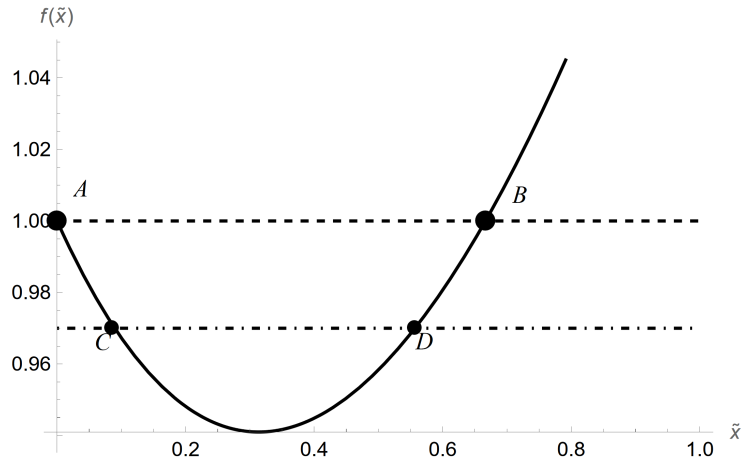

Fig. 3 Plot for the function $f(\tilde{x})$ in Eq. (7) with $r = 5$.

The dashed line intersects with the curve at two points A and B, which shows that there are two values of $x$ giving the same period $T_p$, as is expected. The plot confirms that there is only one minimum for the period $T$, which shows that Occam's razor does work very well in this case. The vanishing of the derivative of $f(\tilde{x})$ with respect to $\tilde{x}$ determines the location of this lone minimum, that is,

$$3\tilde{x}^2 + 3r\tilde{x} - r = 0. \tag{8}$$

The root of our interest is

$$\tilde{x}_m = \frac{1}{2}\left(\sqrt{r\left(r + \frac{4}{3}\right)} - r\right). \tag{9}$$

Substituting $\tilde{x}_m$ into $f(\tilde{x})$ gives its minimal value, that is,

$$f(\tilde{x}_m) = \frac{3}{2}\left(\sqrt{r\left(r + \frac{4}{3}\right)} - r\right) = 3\tilde{x}_m. \tag{10}$$

Thereby, from Eq. (7) the minimum period is obtained as

$$T_m = T_p\sqrt{3\tilde{x}_m}. \tag{11}$$

It is of interest to examine the limit of $T_m$ as $r \to \infty$, that is, when the putty's mass is negligibly small compared with the rod's mass. Hence, we expect to see $T_m$ go to $T_p$ or $3\tilde{x}_m$ go to 1 as $r \to \infty$. For this purpose, we find

$$\lim_{r\to\infty} \tilde{x}_m = \lim_{r\to\infty} \frac{1}{2}\left(\sqrt{r\left(r + \frac{4}{3}\right)} - r\right) = \frac{1}{2}\lim_{r\to\infty}\frac{\frac{4}{3}r}{\sqrt{r\left(r + \frac{4}{3}\right)} + r} = \frac{1}{3}, \tag{12}$$

where the second equality is obtained by using the identity

$$\left(\sqrt{r\left(r + \frac{4}{3}\right)} - r\right)\left(\sqrt{r\left(r + \frac{4}{3}\right)} + r\right) = \frac{4}{3}r. \tag{13}$$

Thereby, from Eq. (11), as $r \to \infty$, $T_m$ does indeed goes to $T_p$ as expected.

## Discussions

The interplay between physics and mathematics has been a recurring theme in my teaching even at the introductory level. Whenever possible, I always encourage students to think hard about physics before jumping into calculation. Only when they have extracted as much physics as possible from thinking and finally squeezed out a question (such as the exact location for minimum period) that cannot be answered without calculation, do we then start calculating and let mathematics perform its duty. For the problem in this article, without calculation, we can only know that there exists at least one minimal period when $0 < x < 2L/3$. However, we cannot obtain the precise value of $x$ that gives the minimal period by pure thinking; careful calculation now becomes inevitable.

Two remarks are in order.

1. The location $x_p = 2L/3$ for putty on the rod that does not change the period of the physical pendulum is actually the center of percussion of the rod. This result is not restricted to the case with uniform rod but is generally true for a rigid body with arbitrary shape [7]. Question (e) in Exercise 1 is designed for this purpose.
2. The existence of a minimal period for $0 < x < 2L/3$ is guaranteed by the so-called Rolle's Theorem: Suppose that $y = f(x)$ is continuous over the closed interval $[a, b]$ and differentiable at every point of its interior $(a, b)$. If $f(a) = f(b)$, then there is at least one number $c$ in $(a, b)$ at which $f'(c) = 0$ [8]. In our context, the horizonal line AB crosses the curve twice in Fig. 1 (a), thereby, there is at least one minimum between A and B whose tangent is horizontal.

Two exercises, two additional remarks, and some backgrounds to my teaching are included in the Supplementary Materials.[9]

## Acknowledgments

The author is grateful to anonymous referees for constructive suggestions to improve the paper quality. He also thanks Nainai for encouraging him to publish.